\documentclass[11pt]{article}

\usepackage[margin=1in]{geometry}
\usepackage{amsmath}
\usepackage{booktabs}
\usepackage{graphicx}
\usepackage[hidelinks]{hyperref}
\usepackage{microtype}
\usepackage{parskip}
\usepackage{xcolor}

\definecolor{linkblue}{HTML}{24527A}
\hypersetup{
  colorlinks=true,
  linkcolor=linkblue,
  citecolor=linkblue,
  urlcolor=linkblue
}
\title{\textbf{Self-Driving Negotiator}\\
\vspace{5pt}
\large An interactive, verifiable benchmark for social negotiation and theory of mind under hidden intent, built on Prime Intellect's \texttt{verifiers}.}
\author{\textbf{Ashutosh Kumar} \\ Machine Learning Engineer, Owl Autonomous Imaging, Inc.\thanks{Work done outside of Owl AI}}

\begin{document}
\maketitle

\begin{abstract}
Autonomous driving is full of tiny social negotiations: a driver presses forward, another yields, a pedestrian fakes toward the curb, or a lane vehicle chooses whether to open a merge gap. Such interactions require inferring hidden intent from behavior under partial observability and then acting safely and efficiently. Existing autonomous-driving language benchmarks mostly focus on perception, visual question answering, or open-loop planning, while existing language-agent negotiation benchmarks typically make the negotiation explicit in text. \texttt{Self-Driving Negotiator} bridges the gap between the two: a text-only, multi-turn, procedurally generated environment for measuring implicit social coordination in driving. Agents generate specific driving actions. Reward and diagnostics are computed from the privileged simulator state, not from the explanation of the model. This report covers task design, reward and anti-gaming invariants, validated scenarios, non-LLM baselines, and a six-model inference leaderboard. Current models are far removed from the scripted expert. The best average success rate across three scenarios is 0.68; \texttt{contested merge} is statistically flat across models; and difficulty tiers separate cue-following from true wait-for-commitment behavior. The code is available on Prime Intellect's Environment Hub\footnote{\url{https://app.primeintellect.ai/dashboard/environments/ashu1069/self-driving-negotiator}}.
\end{abstract}

\section{TL;DR}
Imagine two self-driving cars reaching an unmarked intersection simultaneously (a frequent occurrence on suburban roads). There are no traffic signals, no indicators, and no communication (self-driving cars are unable to converse with one another verbally, regrettably). At this moment, one of the vehicles must proceed first, and if they both move simultaneously, a collision will occur. \textit{What method can they use to determine this?} A human operator observes the actions of the other driver: \textit{are they reducing speed (allowing you to merge) or maintaining their pace (asserting their right of way) and deciding to overtake?} We deduce the other individual’s intentions based on their actions, then respond accordingly. People perform this activity countless times each day. The issue represents a minor, specific example of the theory of mind~\cite{frith2005theory}, representing another agent's concealed state while facing time constraints and safety risks.

The human ability to negotiate while driving is transformed into a quantifiable benchmark for language models by \texttt{Self-Driving Negotiator}. Every turn, the agent receives a text description of the traffic situation and outputs a driving decision (continue, yield, stop, plus acceleration). The agent is never informed of the other road users' \textit{hidden disposition}, which must be deduced from their actions alone. A verifiable, non-gameable reward calculated from the privileged simulator state, rather than whatever the models say, is used to measure performance. A Prime Intellect \texttt{verifiers} multi-turn environment is used to implement it.

\section{But why even bother?}
Let me begin by outlining the limitations of this work: the scene is perceived as a textual description, and LLM agents reason over the description. There is no perception involved, and future work will depend on working with Vision Language Models (VLMs) / Vision Language Action (VLA) models, but that is not the point. Instead, we will talk about a reasoning paradigm related to these self-driving scenarios and see if the LLMs can determine this through their chain of reasoning without visual support. Visual Question Answering (VQA), one of the perception-oriented autonomous-driving benchmarks, asks models to describe a scene, respond to inquiries, or generate plans based on observations~\cite{drivelm}. Although they are not LLM-native text environments, driving simulators like Nocturne~\cite{nocturne} and Waymax~\cite{waymax} facilitate interactive multi-agent research and usually focus on imitation learning or reinforcement learning agents. Separately, language-agent benchmarks like $\tau$-bench~\cite{taubench} examine verifiable results and dynamic interactions, but their domains are tool-use discussions rather than embodied, safety-grounded coordination. None of them measures social negotiation skills, which a human/agent needs while driving. More details on the comparisons between related benchmarks can be found in Table~\ref{tab:benchmark_families}
\begin{table}[t]
    \centering
    \small
    \setlength{\tabcolsep}{4pt}
    \renewcommand{\arraystretch}{1.15}
    \begin{tabular}{
        p{0.19\linewidth}
        p{0.2\linewidth}
        p{0.28\linewidth}
        p{0.25\linewidth}
    }
    \toprule
    \textbf{Family} &
    \textbf{Examples} &
    \textbf{What they test} &
    \textbf{What they do not test} \\
    \midrule

    \textbf{Perception / VQA driving} &
    DriveLM~\cite{drivelm}, NuPlanQA~\cite{park2025nuplanqa}, Bench2ADVLM~\cite{zhang2025bench2advlm}, LaMPilot~\cite{ma2024lampilot} &
    Scene understanding and open-loop decision queries, e.g., ``What is in this scene?'' or ``What should the ego vehicle do?'' These benchmarks primarily test single-agent perception, captioning, and open-loop planning. &
    The \textit{interactive} component: how other agents react to the ego vehicle's behavior. \\

    \textbf{LLM negotiation / game theory} &
    LLM-Deliberation~\cite{abdelnabi2023llm}, LLMsPark~\cite{chen2025llmspark}, NegotiationArena~\cite{bianchi2024well} &
    Bargaining, negotiation, and matrix-game reasoning with explicit, often text-mediated offers. &
    \textit{Embodied, real-time, safety-grounded} coordination under \textit{implicit} signaling. \\

    \bottomrule
    \end{tabular}
    \caption{Comparison between related benchmark families and the interactional capabilities they leave under-tested.}
    \label{tab:benchmark_families}
\end{table}

The skill this work is focused on: \textbf{infer another agent's latent intent from its behavior and then coordinating in real time without explicit communication}, is exactly what safe driving (and a lot of multi-agent autonomy) requires, and it is underserved. It is also a clean, bounded testbed for \textit{theory of mind}, a capability the field increasingly cares about but rarely measures in a way that is hard to game. A second methodological gap exists. A lot of benchmarks are \textit{static} (datasets): a fixed gold answer, a fixed question. Because the appropriate course of action relies on how the other agent reacts to what you just accomplished, the driving negotiations cannot be depicted in a static scenario. Let's say I successfully compress the theory of mind to one-step classification by freezing it into a Q\&A benchmark. A negotiation benchmark needs to be an environment rather than a dataset, particularly in self-driving scenarios.

In conclusion, I assert that a small, sharp, interactive environment is a more robust probe of agentic social reasoning than either a perception VQA set or an explicit bargaining game if it (1) isolates the hidden-intent interface, (2) scores it with a reward that cannot be faked by talking, and (3) resists saturation.
\section{Let's take a measurement (task thesis)}
\label{sec:measurement}

\begin{figure}[hbt!]
    \centering
    \includegraphics[width=\linewidth]{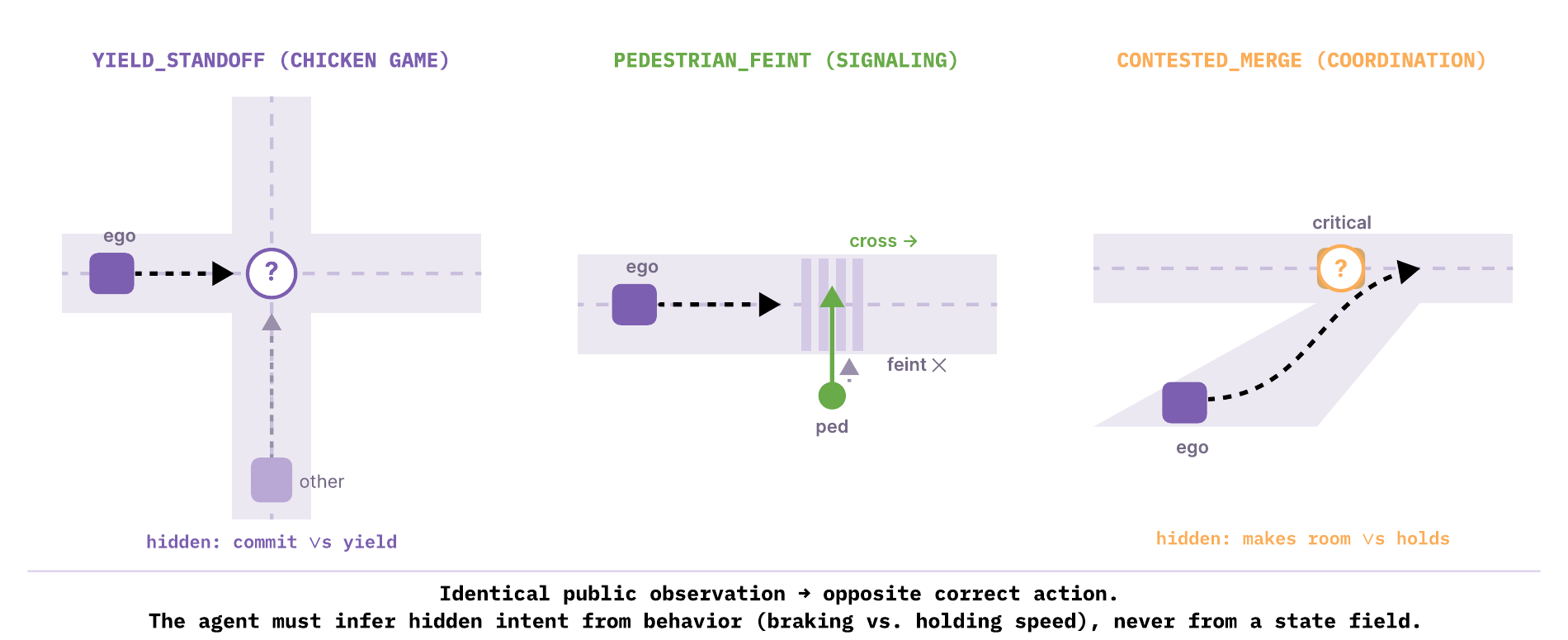}
    \caption{Three validated negotiation scenarios, each an instance of inferring a hidden disposition from behavior. In every scenario the opposing agent holds a per-episode intent that is never placed in the observation; the agent sees only its consequences (braking vs. holding speed). The correct action is opposite under the two intents, so no constant policy can win, the defining invariant of the benchmark}
    \label{fig:scenarios}
\end{figure}

The environment evaluates a model's (agent's) ability to use implicit social signaling to negotiate right-of-way. Under \textit{partial observability} (world model guys can hold their breath) and without explicit rules, the model must infer the latent disposition of another road user from behavior and select proceed/yield/stop actions over multiple turns. Success is determined by safe, efficient, verifiable coordination. Every round, the agent is given a natural-language traffic scene and is required to provide a JSON action:

\begin{verbatim}
{"reasoning": "...", "acceleration": 0.5, "steering": 0.0, "maneuver": "proceed"}
\end{verbatim}

The hidden state is never exposed in the public observation. The model only sees positions, speeds, recent motion, and previous messages. The reward is computed from the simulator state after the episode: whether the ego vehicle reached its goal, collided, stalled, acted efficiently, and selected maneuvers compatible with the
opponent's latent disposition. Three scenarios are validated instances of the same underlying tasks as shown in Figure~\ref{fig:scenarios}.

In all of the above-mentioned validated scenarios, no single action is successful; instead, the agent must "read" (LLM-based, literally reading-only) the environment and adapt.  Consequently, the main argument is that no action is universally correct. The right course of action is determined by a latent inclination that must be inferred over time. Because the item being evaluated would be eliminated if the challenge were restricted to single-turn question answering, the benchmark becomes an environment rather than a static dataset.

\section{Reaching the intuitive core (design principles)}
\label{sec:design}
The benchmark design's goal is to ensure that it measures the correct variable while resisting the inappropriate shortcut. To reach the intuitive core of this benchmark, let's go through the following five design principles.
\subsection{The agent should infer, not get told (hidden disposition + behavioral cue)}

Each episode begins with the opposite agent sampling and holding a disposition, such as committed versus yielding. \textit{It's crucial to remember that this is never included in the observation}. Only behavior (position, speed, and whether the other agent is braking) is visible to the agent. The agent model cannot view the disposition or examine the concealed state (but a grader or a public observer can). This distinction is crucial because if the concealed state seeps into the prompt, a social inference benchmark collapses into a search or a lookup problem.

\subsection{Interaction is the whole point (environment, not a dataset)}
The "other" agent on the path, or the opponent, responds to the ego. A yielder only shows up when you get close; otherwise, it may just appear to be a committed car that keeps coming. For instance, turns 1-2 determine the right move at turn 3. This cannot be frozen into a static dataset (observation $\rightarrow$ label) without destroying the object being monitored.

\subsection{Score behavior, not the prose (verifiable + un-gameable)}

A clean goal completion is always best (\textbf{+1}, with minimal efficiency/comfort shaping), a collision is always worse (\textbf{-1}), and partial progress falls in between. The scored reward is constrained and clearly aligned with the success. Keyword-style reasoning bonuses were included in an earlier version of the prototypes, but they were eliminated since they rewarded a model for using terms like ``safe" and ``yield" (which may be easily gamed, and we're interested in the agent's behavior, not the writing). Instead, I included a metric called \texttt{decision\_correctness} that was calculated by comparing the agent's actual moves with the \textit{privileged} concealed state (disposition). A model that provides a convincing explanation or uses the appropriate language but acts incorrectly receives a score of \textbf{0}, while a model that says nothing but yields to a committed car scores. The metric measures what the agent did, not what it claimed (sounds trivial at the end, anyways).

\subsection{Decide less often than you steer (separate decision and control)}

While the LLM only acts per \texttt{decision\_interval} steps, physics advances at a brief simulator step. This keeps evaluation within a realistic turn budget and reflects the division between planning and low-level control in autonomous-driving stacks. The LLM determines each \texttt{decision\_interval} step (by default, 5 $\rightarrow$ 0.5 s), while physics operates at 0.1 s stages. This effectively allows an episode to conclude within a reasonable LLM turn budget rather than requiring hundreds of calls, and it mimics real AV stacks (a planner re-plans significantly less frequently than the controller actuates).

\subsection{``Difficulty" = information structure, not noise}

Finally, the project's most significant design lesson is that observation noise is not the crucial difficulty lever. Since multi-turn models just average the noise over turns, my initial attempt at a difficulty knob, which added \textbf{per-step Gaussian noise} to the behavior cue, was totally ineffective against them (DeepSeek scored 20/20/20 across ``harder" tiers; we shall see more specifics later). Making it difficult to determine what information is available and when was the solution. The information structure is altered in the following ways by the current design:

\begin{itemize}
  \item \textbf{Delayed reveal:} Both dispositions look identical (a neutral cruise) until the ego vehicle/agent is close, and then they diverge. The information is \textit{absent} early, so memory cannot recover it. The correct play here is to hedge (approach slowly and be ready to stop) until intent resolves.
  \item \textbf{Bluff / feint:} The opposition first shows the opposite of its true intent (a committed car soft-brakes to look like it is yielding) and reverses the instant you take the bait. A single instantaneous cue is not unsafe, and we must wait for a \textit{committed} signal (the other actually stopping). This is a genuine theory of mind stressor, and unlike noise, it actually bites strong models in Section
\end{itemize}

\begin{figure}[hbt!]
    \centering
    \includegraphics[width=\linewidth]{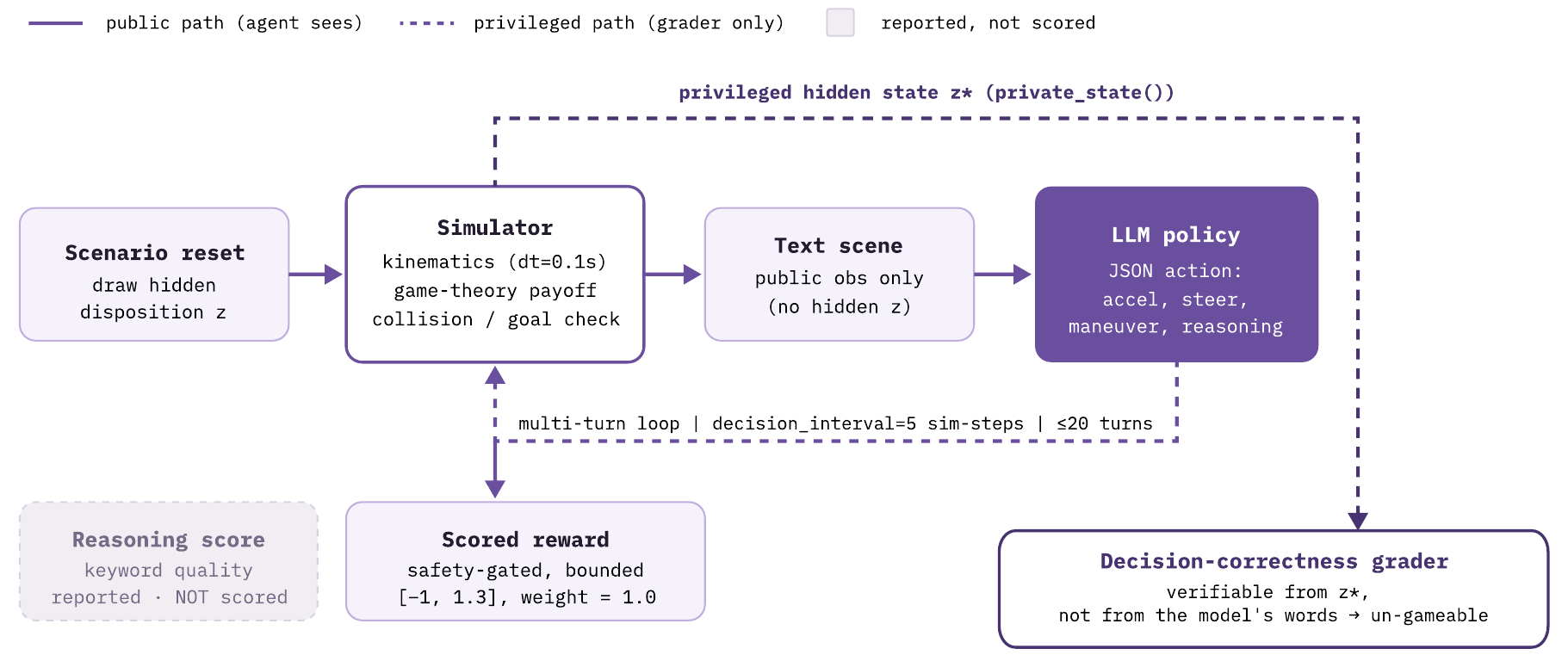}
    \caption{The benchmark separates what the agent sees from what the grader knows. The simulator draws a hidden disposition each episode and exposes only a public text scene; the LLM acts over a multi-turn loop. Two graders run in parallel: a bounded, safety-gated scored reward and a decision-correctness check computed from privileged ground truth, so neither keyword-stuffing the reasoning field nor a constant policy can inflate it. Reasoning-keyword quality is reported but deliberately excluded from the reward.}
    \label{fig:env_eval_design}
\end{figure}

Conclusively, these tiers reward hedging until the other agent's commitment becomes
observable, rather than overreacting to a single cue. Aggregating the design decisions, the complete environment and evaluation pipeline design can be visualized in Figure~\ref{fig:env_eval_design}.
\section{The social negotiation scenarios while driving}
We have already introduced the three validated scenarios in Section~\ref{sec:measurement}. It's time to discuss them in a little more detail, and they map to game theoretic scenarios. All three are \textbf{text-only} (any LLM can play, no vision required), procedurally generated (geometry, timing, and the hidden disposition are sampled per episode), and seeded for reproducibility.

\begin{itemize}
    \item \texttt{yield\_standoff} (chicken game): Heard of the chicken game~\cite{osborne1994course}? Essentially, it's a model of conflict between two players in game theory. In our case, we have a T-intersection, two cars. The other car is \textit{committed} (drives through, and we must yield) or a \textit{yielder} (brakes to let you go). The optimal play here is to \textit{read the room}, yield to the committed, and take the gap from the yielder. A constant \texttt{PROCEED} collides with committed cars, and a constant \texttt{STOP} puts us in a deadlock situation.
    \item \texttt{pedestrian\_feint} (signaling game): A signaling game~\cite{fudenberg2026game} is a type of dynamic Bayesian game in which one player takes action, the signal, to convey information to another player. Here, the two players are a car and a pedestrian at the light. A pedestrian either genuinely crosses or feints (lurches toward the road, then stops short). The car must brake for real crossers and not overreact to feints. The benchmark penalizes both recklessness (hitting a crosser) and over-caution (stalling forever). This is a classic case we discussed in the design~\ref{sec:design} as \textbf{delayed-step-out}, i.e., later commitment = less reaction time.
    \item \texttt{contested\_merge} (coordination game): A coordination game~\cite{cooper1999coordination} describes the situation where a player will earn a higher payoff when they select the same course of action as another player. In our case, this game has been mapped onto a longitudinal on-ramp merge situation in driving (contested merge). The agent controls speed to slot into the gap around one lane vehicle with a hidden ``make room" disposition. It is reformulated as a 1-D speed decision (lateral position auto-derives from progress, as in real merge assist), so the decision is clean. A constant \texttt{PROCEED} side-swipes committed vehicles, and a constant \texttt{STOP} never merges.
\end{itemize}

I did consider a lot of other scenarios that require social negotiation skills, such as \texttt{four\_way\_stop}, \texttt{parking\_lot}, \texttt{uncertainty\_reasoning}, and \texttt{driving\_theory\_verifier}, but later removed them because they were either gameable or trivially graded, and shipping them would compromise the integrity of this benchmark. Find the mechanistic details of the three validated scenarios in Table~\ref{tab:scenarios}.
\begin{table}[t]
\centering
\small
\resizebox{\linewidth}{!}{%
\begin{tabular}{p{0.22\linewidth}p{0.27\linewidth}p{0.25\linewidth}p{0.18\linewidth}}
\toprule
Scenario & Hidden state & Observable cue & Naive failure \\
\midrule
\texttt{yield\_standoff} &
Whether the other car will commit or yield at the intersection &
Braking, holding speed, or waiting near the conflict point &
Always proceed collides; always stop deadlocks \\
\texttt{pedestrian\_feint} &
Whether the pedestrian is truly crossing or only feinting &
Step into lane versus stop short &
Proceed hits crossers; stop never clears \\
\texttt{contested\_merge} &
Whether the lane vehicle will make room &
Opening or closing the longitudinal gap &
Proceed sideswipes; stop never merges \\
\bottomrule
\end{tabular}}
\caption{Validated scenarios: each requires inference from behavior rather than a fixed action.}
\label{tab:scenarios}
\end{table}

\begin{figure}[hbt!]
    \centering
    \includegraphics[width=\linewidth]{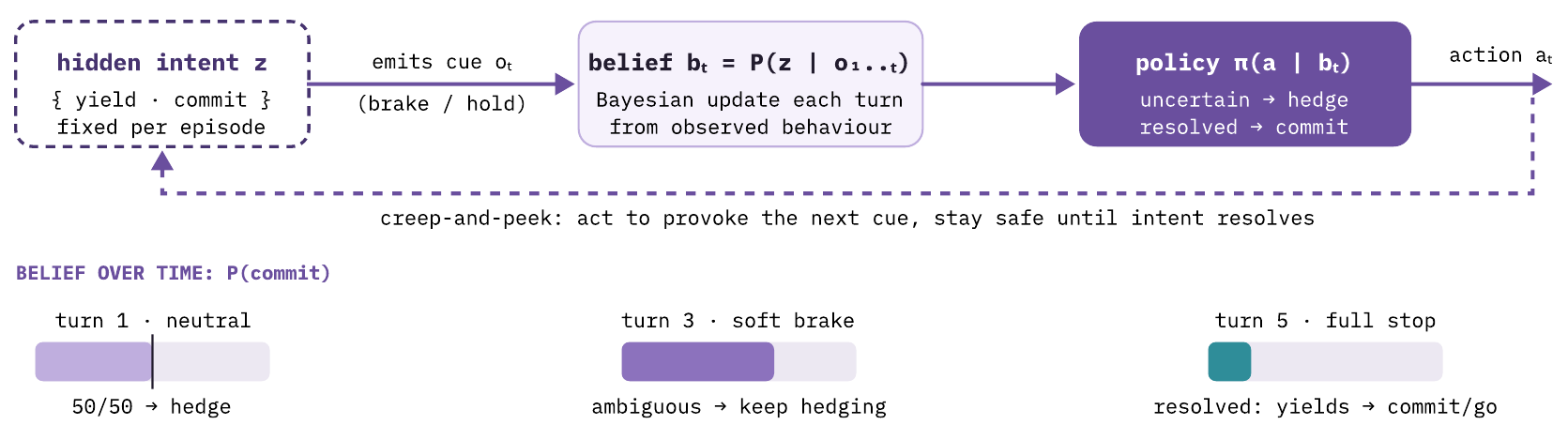}
    \caption{\textbf{Every scenario is a small POMDP: act well only by tracking a belief over the other agent's hidden intent}. Because intent is never observed (and, at higher tiers, is revealed late or bluffed), a single cue is unsafe to act on. The competent policy maintains a belief, keeps it alive by creeping to provoke informative cues, and commits only once the belief resolves, which is exactly the reference expert's behavior and what most LLMs fail to do.}
    \label{fig:belief_state}
\end{figure}

I want to talk about another way to frame the scenarios, where each one is a small, partially observed Markov decision process (POMDP). This makes it possible to think of the tasks as belief-state tracking under hidden intent (the world modeling guys must be smacking their lips). It is depicted in Figure~\ref{fig:belief_state}.
\section{Keeping the benchmark honest (a tough nut)}
A benchmark is only as good as its resistance to ``wrong" shortcuts (or shortcuts?). This project encodes ``what it means to be a valid negotiation task" as \textbf{executable invariants}, and runs it on every commit. A scenario is
validated only if it satisfies invariants that encode what a negotiation task
must be:

\begin{itemize}
  \item \textbf{Anti-gaming}: The scripted \textit{expert must strictly beat every constant policy} such as \texttt{always\_proceed} and \texttt{always\_stop}. If a fixed action matched the expert, the task would not be measuring negotiation. A set of examples are shown in Figure
  \item \textbf{Solvable}: The task must be solvable by the expert within the LLM turn budget. Numerically, a reference expert reaches $\geq 0.7$ success, meaning the task is not impossible.
  \item \textbf{Hidden disposition}: The ground-truth disposition never leaks into the public observation and is only visible to the grader.
  \item \textbf{Reward ordering}: A clean goal must score above a collision.
  \item Difficulty tiers must remain solvable while punishing naive cue-following.

\end{itemize}
\begin{figure}[hbt!]
    \centering
    \includegraphics[width=\linewidth]{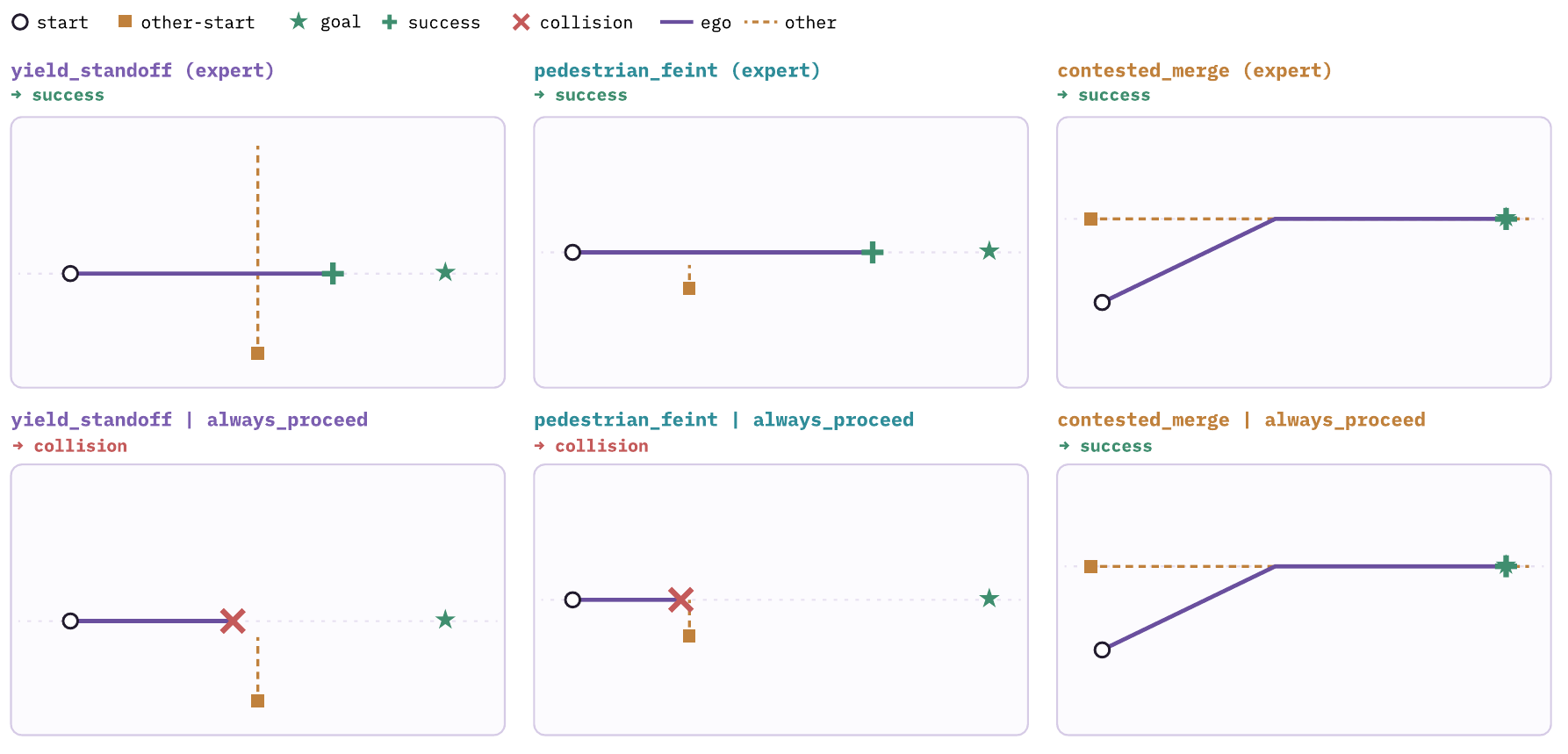}
    \caption{\textbf{The invariant, made visual: inferring intent succeeds where a constant policy crashes}. Top-down ego paths (purple) and the other agent (dashed clay) for the behavior-inferring expert (top row) and the constant \texttt{always\_proceed} policy (bottom row). The expert creeps-and-yields to clear all three scenarios; constant \texttt{PROCEED} collides at the intersection and the crosswalk ($\textbf{X}$), and only survives the merge, where simply going happens to coincide with the gap, which is exactly why \texttt{contested\_merge} sits at the guessing floor }
    \label{fig:placeholder}
\end{figure}
These tests prevent common benchmark failures: leakage, trivial constant-action
solutions, impossible scenarios, and reward inversions.

\section{Evaluation Protocol and Baselines}
\label{sec:evaluation_protocol}

I evaluate each policy in a multi-turn interactive setting implemented as a Prime Intellect \texttt{verifiers} \texttt{vf.MultiTurnEnv}. At every turn, the model receives a system prompt and a natural-language scene description, then returns a JSON action of the form
\texttt{\{reasoning, acceleration, steering, maneuver\}}. The episode continues until success, collision, timeout, or another terminal condition. Since LLMs are creative machines even when asked to be boring, parsing is deliberately protective: malformed JSON, prose inside numeric fields, or invalid actions degrade to conservative defaults rather than crashing the evaluation. In other words, bad formatting is punished as bad driving, not rewarded with a stack trace.

\begin{figure}[hbt!]
    \centering
    \includegraphics[width=\linewidth]{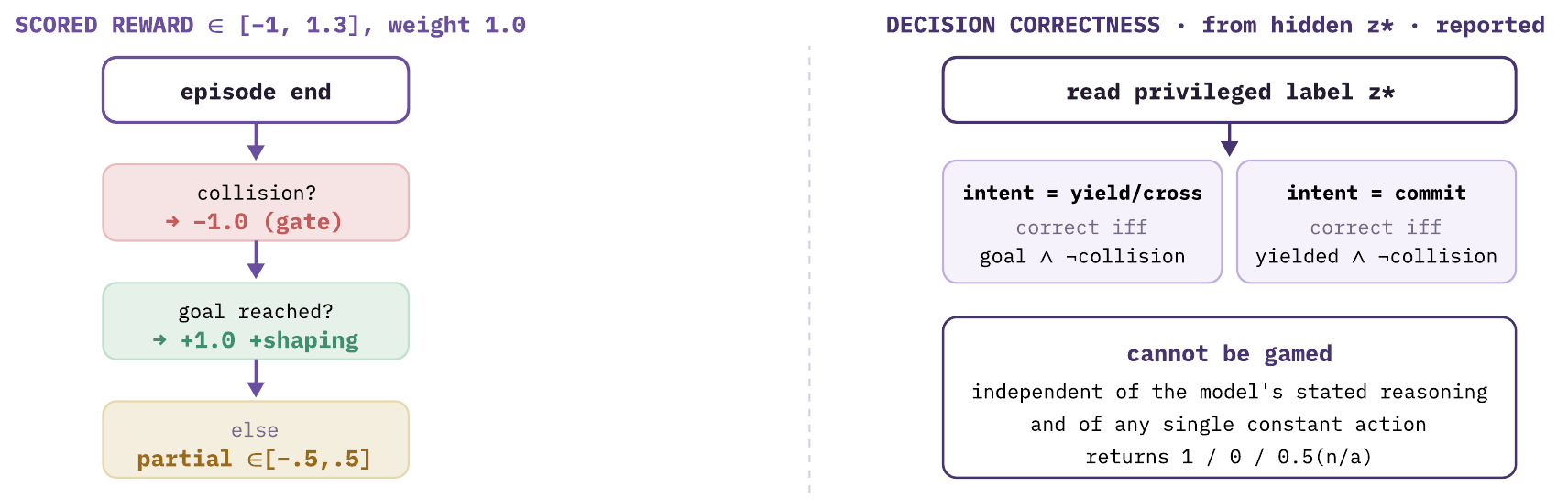}
    \caption{\textbf{Two graders, two purposes: a bounded scored reward and an ungameable correctness check}. The scored reward is lexicographic: collision dominates negatively, a clean goal is best, and partial progress lies in between. Decision correctness is evaluated against the hidden disposition $z^\ast$, so it certifies why an episode succeeded rather than trusting the model's words.}
    \label{fig:metrics}
\end{figure}

The headline metric is success rate over seeded episodes. We additionally report mean episode length, collision rate, bounded outcome reward, \texttt{decision\_correctness}, and per-dimension diagnostics for safety, social compliance, efficiency, and comfort. The scored reward is computed by \texttt{compute\_episode\_reward}; it is bounded and success-aligned, with collision dominating negatively, clean goal completion rewarded positively, and partial progress placed strictly between. The second grader, \texttt{decision\_correctness}, is intentionally not a reward-shaping term. It evaluates whether the policy selected the correct maneuver relative to the hidden disposition $z^\ast$, making it a verifiable theory-of-mind signal rather than a trust-me-bro explanation from the model. Figure~\ref{fig:metrics} summarizes the distinction between these two graders.

For LLM policies, confidence intervals are percentile bootstrap 95\% intervals computed from stored per-rollout outcomes. For the non-LLM baseline table, we report normal approximations. Unless otherwise stated, all headline runs use $n=200$ seeded episodes per scenario and policy, which is large enough for the intervals to be interpretable without pretending the environment is a physics experiment.

Non-LLM policies anchor the scale. The reference expert is not a hidden superhuman driver; it is a small set of behavior-conditioned rules, such as creep-and-peek at an intersection or slowing to merge behind a committed lane vehicle. Constant policies deliberately ignore the latent state: \texttt{always\_proceed}, \texttt{always\_stop}, \texttt{cautious}, and \texttt{random}. This gives the evaluation an important sanity check: a model that does not beat \texttt{always\_proceed} is not negotiating; it is just confidently guessing in JSON.

\begin{figure}[hbt!]
    \centering
    \includegraphics[width=\linewidth]{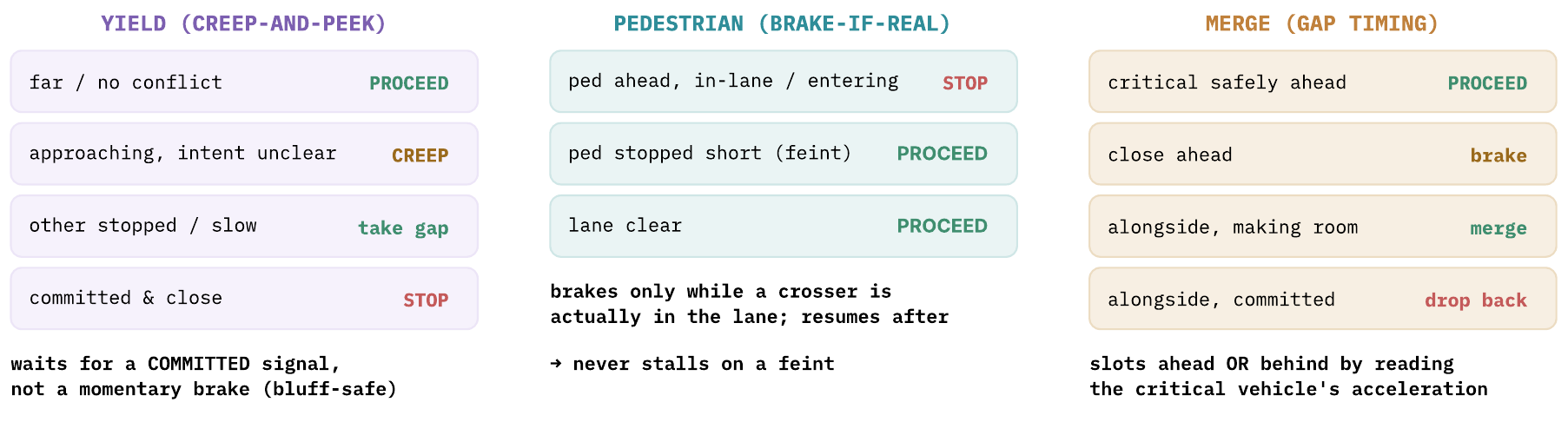}
    \caption{\textbf{What the reference expert actually does}. The expert is a compact set of behavior-conditioned rules; the LLM's job is to rediscover the same dependency from text. Each scripted policy maps an inferred situation to a maneuver, defining the empirical ceiling. The point is not that these rules are sophisticated, but that they condition on the other agent's behavior, which is precisely the dependency constant policies lack and many LLMs underuse.}
    \label{fig:expert}
\end{figure}

\begin{figure}[hbt!]
    \centering
    \includegraphics[width=\linewidth]{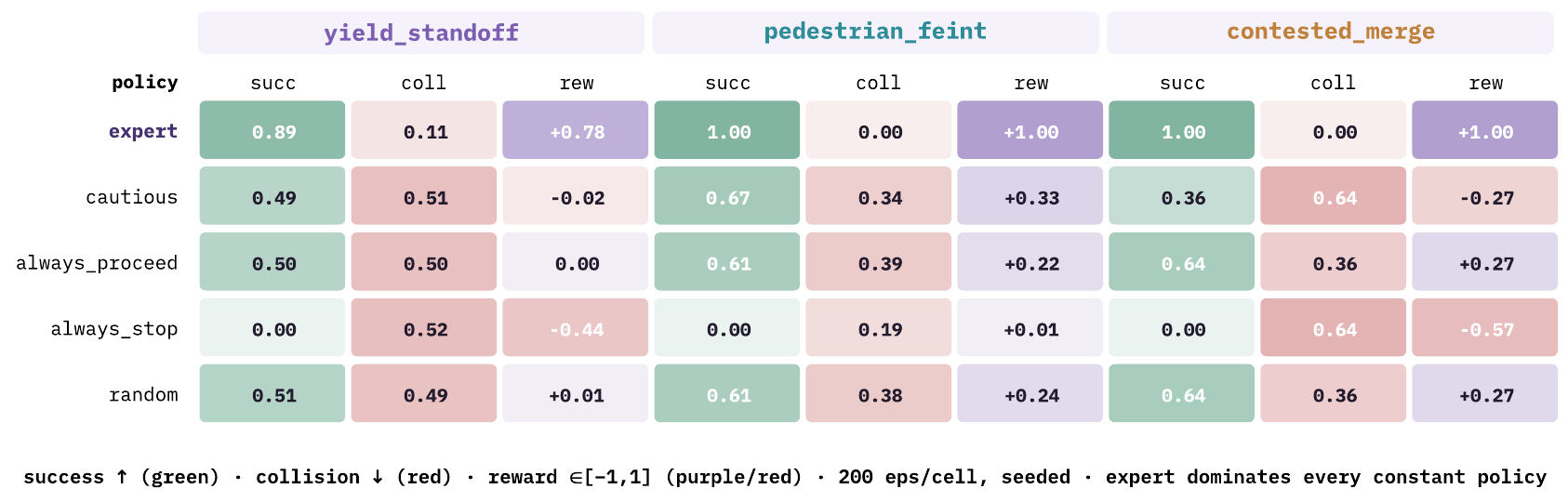}
    \caption{\textbf{The shipping invariant: a behavior-inferring expert beats every constant policy, everywhere}. Full non-LLM baseline grid - success, collision, and bounded outcome reward across all three scenarios and five reference policies. The expert row dominates; \texttt{always\_stop} never reaches a goal; \texttt{always\_proceed} and random coincide (the simulator is largely deterministic given the seed).}
    \label{fig:baselines}
\end{figure}

Figure~\ref{fig:baselines} shows the success rates of the expert and constant non-LLM policies. The key property is the gap between behavior-conditioned experts and state-ignorant constants. If a constant action matched the expert across these scenarios, the benchmark would not be measuring negotiation; it would be measuring whether the safest answer is always ``go'' or always ``stop.'' The observed gap confirms that success requires conditioning on the other agent's behavior, which is the core interaction we intend the benchmark to test. Finally, reproducibility is handled through procedural generation and explicit seeds. Runtime defaults are stored in a Python module, \texttt{config\_defaults.py}, and packaged into the wheel so that a fresh Hub install behaves identically. 

\section{How did the models perform?}
\label{sec:results}

\begin{figure}[hbt!]
    \centering
    \includegraphics[width=\linewidth]{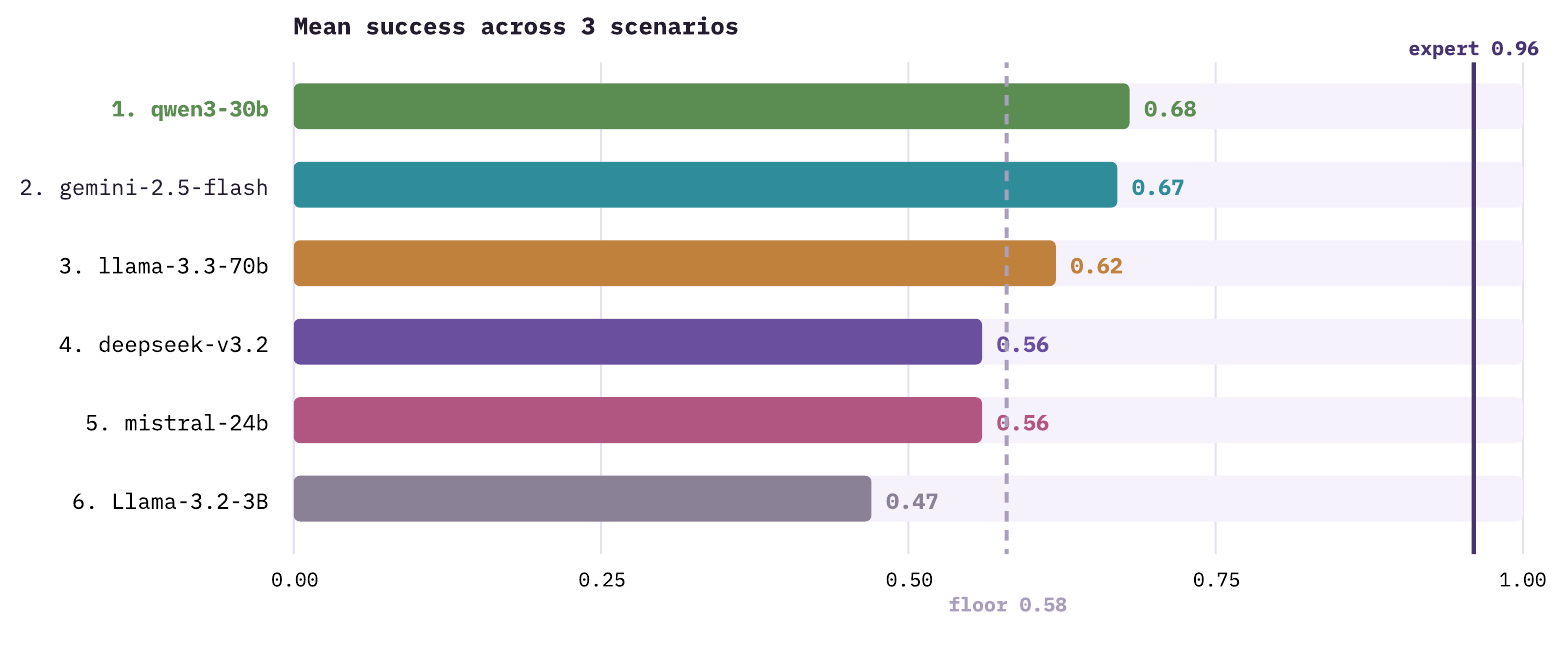}
    \caption{\textbf{Ranked mean}. Qwen (0.68) and Gemini (0.67) overlap and are a top tie; DeepSeek is near-perfect on yield (0.98) yet worst on pedestrian (0.17), so a high mean can hide a brittle profile. LLM and baseline rows are separate 200-sample runs, so floor/ceiling are reference levels rather than paired deltas}
    \label{fig:ranked_mean}
\end{figure}

Table~\ref{tab:llm_results} reports the main success-rate results across the three negotiation scenarios. The best average performance is obtained by \texttt{qwen3-30b-a3b-instruct}, with an average success rate of 0.68, followed closely by \texttt{gemini-2.5-flash} at 0.67 and \texttt{llama-3.3-70b-instruct} at 0.62, as shown in Figure~\ref{fig:ranked_mean}. These models improve over trivial constant behavior in some settings, but none approach the scripted expert, which succeeds in all scenarios. This is the desired regime for a new benchmark: \textbf{the task is learnable, but not saturated}.

\begin{table}[t]
\centering
\small
\begin{tabular}{lcp{2.5cm}p{2.5cm}p{2.5cm}c}
\toprule
 Model & Avg. & \texttt{yield\_standoff} & \texttt{pedestrian\_feint} & \texttt{contested\_merge} & Yield DC \\
\midrule
\texttt{qwen3-30b-a3b-instruct} & 0.68 & 0.83 (0.78--0.88) & 0.71 (0.65--0.77) & 0.50 (0.43--0.57) & 0.83 \\
\texttt{gemini-2.5-flash} & 0.67 & 0.96 (0.93--0.98) & 0.49 (0.42--0.56) & 0.56 (0.49--0.63) & 0.96 \\
 \texttt{llama-3.3-70b-instruct} & 0.62 & 0.92 (0.88--0.95) & 0.44 (0.37--0.51) & 0.50 (0.43--0.57) & 0.98 \\
 \texttt{deepseek-v3.2} & 0.56 & 0.98 (0.96--1.00) & 0.17 (0.12--0.22) & 0.52 (0.45--0.58) & 0.99 \\
\texttt{mistral-small-3.2-24b} & 0.56 & 0.58 (0.52--0.66) & 0.58 (0.52--0.65) & 0.50 (0.43--0.57) & 0.92 \\
\texttt{Llama-3.2-3B-Instruct} & 0.47 & 0.41 (0.34--0.47) & 0.62 (0.56--0.69) & 0.39 (0.33--0.46) & 0.41 \\
\midrule
 Expert reference & 1.00 & 1.00 & 1.00 & 1.00 & -- \\
\texttt{always\_proceed} floor & 0.58 & 0.49 & 0.61 & 0.64 & -- \\
\bottomrule
\end{tabular}
\caption{Success rates for LLM policies across the three negotiation scenarios. Parentheses denote 95\% percentile bootstrap confidence intervals over 200 seeded episodes per scenario and model. \texttt{Yield DC} denotes decision-correctness on the \texttt{yield\_standoff} scenario.}
\label{tab:llm_results}
\end{table}

\begin{figure}[hbt!]
    \centering
    \includegraphics[width=\linewidth]{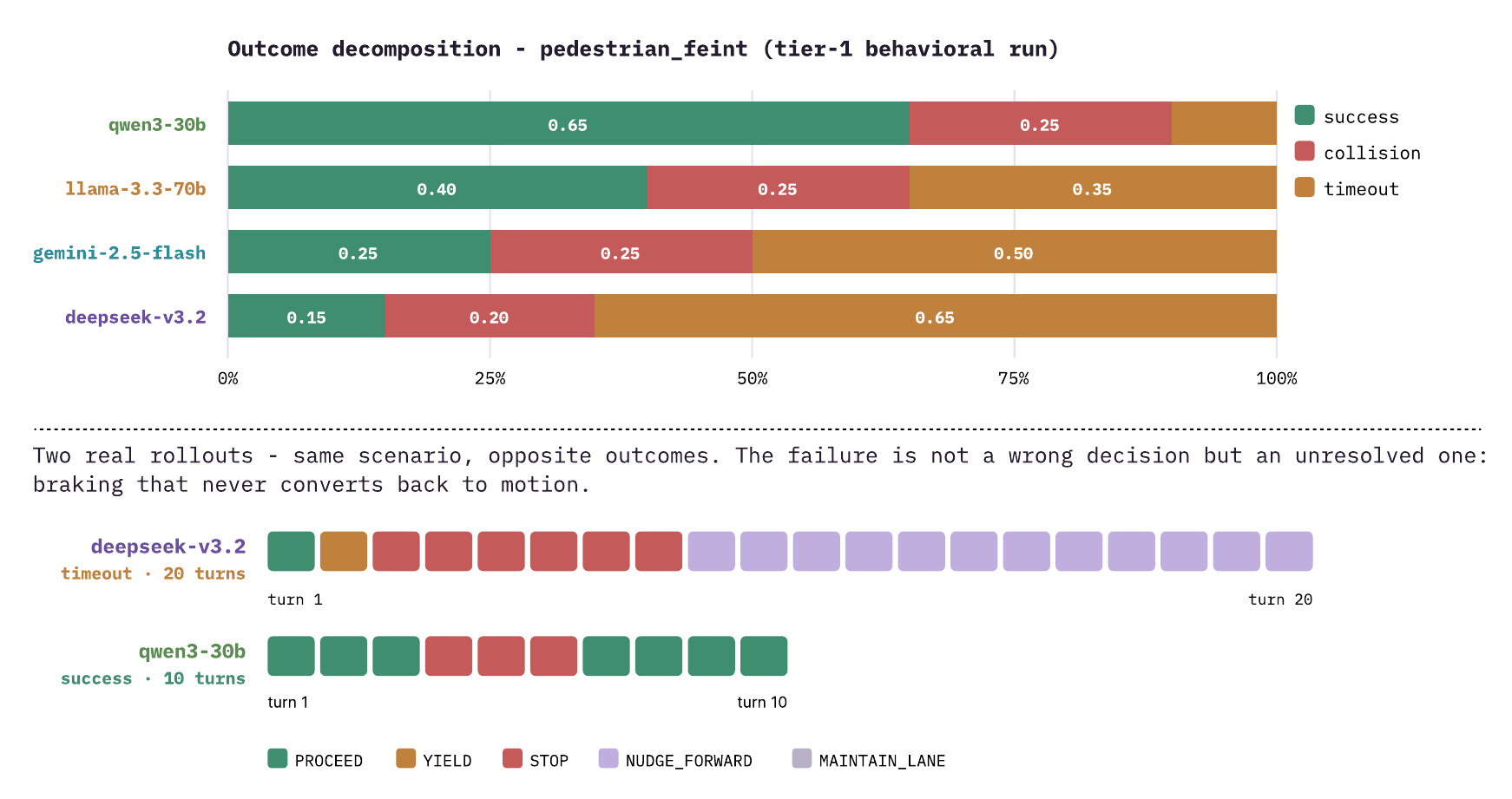}
    \caption{\textbf{On \texttt{pedestrian\_feint}, models fail by over-stopping and stalling, not by hitting pedestrians}. Outcome decomposition (top) shows timeouts, not collisions, dominate the worst models: DeepSeek times out in 65\% of episodes. The maneuver timelines (bottom, real logged rollouts) reveal the mechanism: a productive model (Qwen) brakes only while the crosser is in-lane then commits, whereas DeepSeek brakes for a feint and then \texttt{NUDGE\_FORWARDs} indefinitely without ever clearing. Over-caution is penalized; the benchmark rewards decisiveness once safe.}
    \label{fig:caution_failure}
\end{figure}

The first finding is that caution behaves like a dial, and no model appears to have calibrated it correctly. The strongest intersection negotiators are also among the weakest pedestrian-feint policies: \texttt{deepseek-v3.2}, \texttt{gemini-2.5-flash}, and \texttt{llama-3.3-70b-instruct} achieve 0.98, 0.96, and 0.92 success on \texttt{yield\_standoff}, respectively, but drop to 0.17, 0.49, and 0.44 on \texttt{pedestrian\_feint}. Conversely, the more decisive policies, such as \texttt{qwen3-30b-a3b-instruct} and \texttt{Llama-3.2-3B-Instruct}, are relatively stronger on \texttt{pedestrian\_feint} but weaker on \texttt{yield\_standoff}. This yield-pedestrian inversion suggests that models are not learning a calibrated interaction policy; they are carrying a temperament. Some are too polite, some are too assertive, and the road unfortunately requires both modes on demand. This over-caution failure mode is shown in Figure~\ref{fig:caution_failure}.

\begin{figure}[hbt!]
    \centering
    \includegraphics[width=\linewidth]{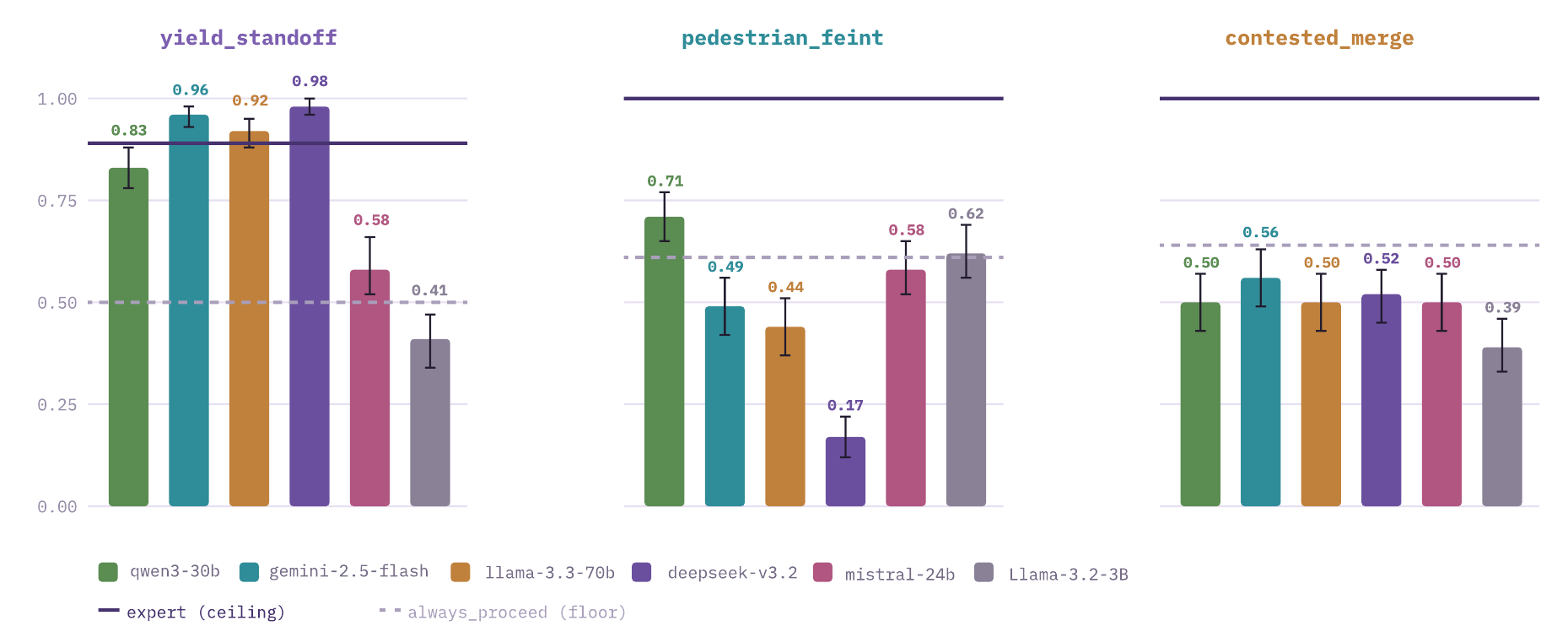}
    \caption{\textbf{No evaluated LLM negotiates better than the scripted expert, and on the merge, every model sits at or below the guessing floor}. Zero-shot success rate for six instruction-tuned models (n=200 rollouts per scenario, tier 1) with bootstrap 95\% CI whiskers, bracketed by two non-LLM references held constant in every panel: the behavior-inferring expert (skill ceiling) and the \texttt{always\_proceed} constant policy (guessing floor). \texttt{yield\_standoff} separates models cleanly and the leaders even edge past the expert reference; \texttt{pedestrian\_feint} is the widest spread, with deepseek collapsing to 0.17 while a small 3B model reaches 0.62; \texttt{contested\_merge} compresses every model into 0.39-0.56, below the 0.64 \texttt{always\_proceed} floor, with overlapping CIs (a statistical tie at, or under, guessing)}
    \label{fig:scenario_results_ci}
\end{figure}

\begin{figure}
    \centering
    \includegraphics[width=\linewidth]{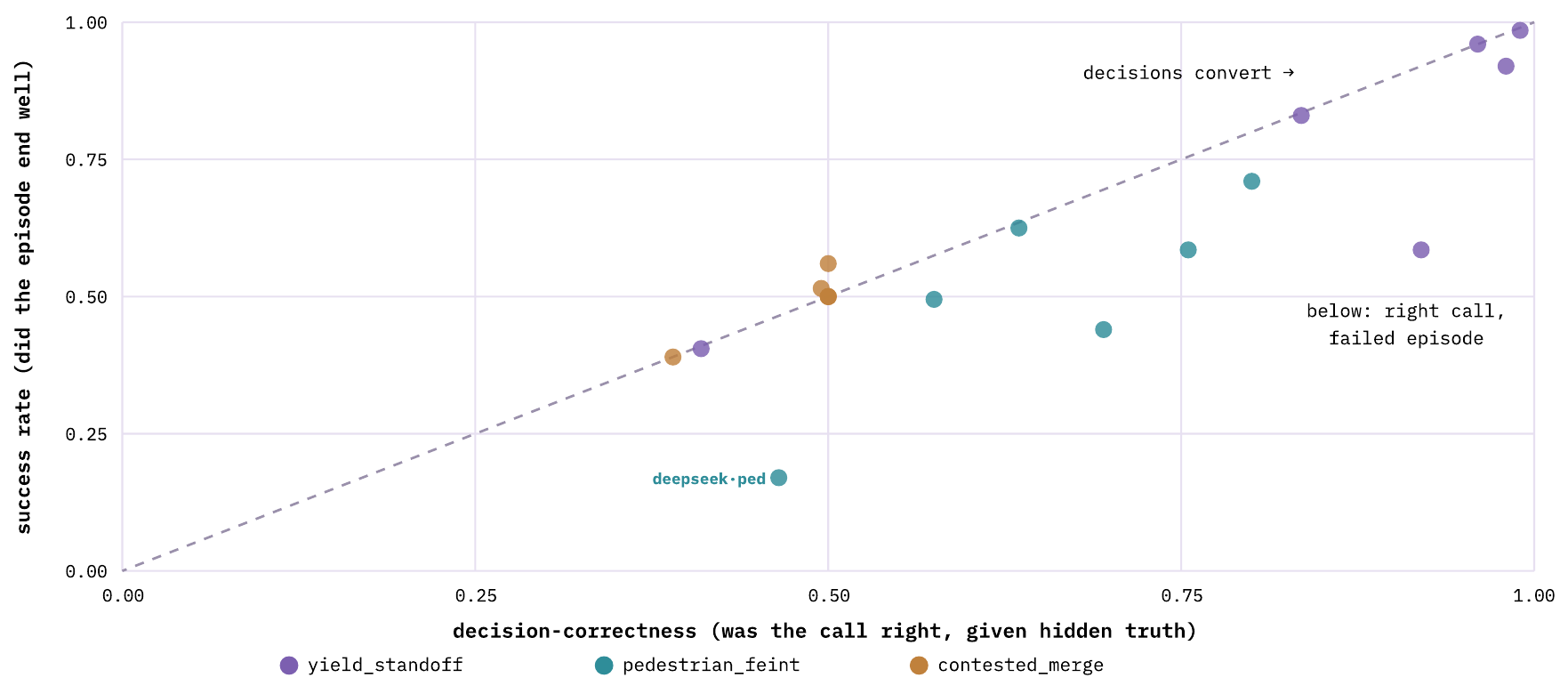}
    \caption{The pedestrian scenario sits below the diagonal: models make the right call yet still fail the episode. Plotting success against decision-correctness, the diagonal marks ``decisions convert to outcomes." Yield points hug it, and merge points cluster at (0.5, 0.5); the entire \texttt{pedestrian\_feint} column falls below it - Deepseek is right 47\% of the time but succeeds only 17\%. The lost distance is over-caution: correct-but-unresolved braking that times out}
    \label{fig:right_wrong}
\end{figure}

The second finding is that recognizing intent is not the same as executing it. This is clearest for \texttt{mistral-small-3.2-24b}: on \texttt{yield\_standoff}, it reaches 0.92 decision-correctness but only 0.58 success. In other words, the model often identifies the committed vehicle correctly but fails to convert that recognition into a completed maneuver. A success-only score would flatten these two cases together. The \texttt{decision\_correctness} grader separates the cognitive part from the control part: reading the other agent, then acting at the right time. The model sees the opening, then somehow still manages to miss the door (shown in Figure~\ref{fig:right_wrong}).

\begin{figure}[hbt!]
    \centering
    \includegraphics[width=\linewidth]{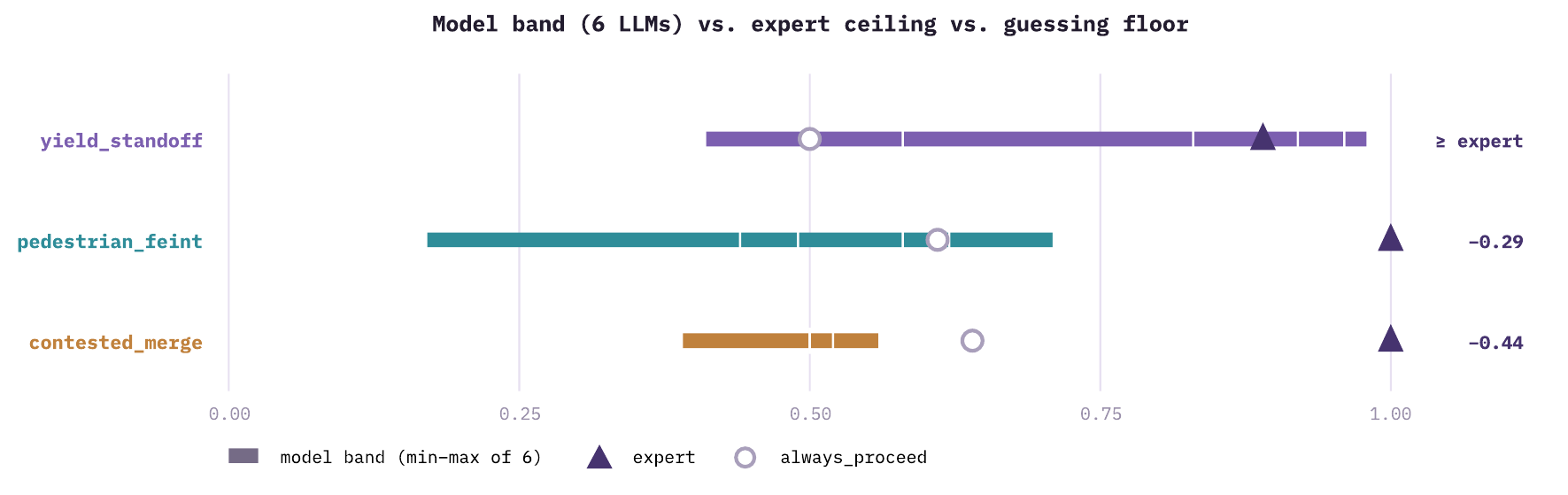}
    \caption{\textbf{The gap to expert is closed on yield but large and uniform on merge}. Each track shows the full spread of the six LLMs (band, with per-model ticks) against the expert ceiling and the \texttt{always\_proceed} floor. On \texttt{yield\_standoff} the best models (up to 0.98) reach or exceed the expert reference; on \texttt{pedestrian\_feint} the band is wide (0.17–0.71); on \texttt{contested\_merge} the entire band (0.39–0.56) collapses left of the floor marker, a 0.44 shortfall to expert and the clearest open problem in the suite.}
    \label{fig:merge_issue}
\end{figure}

The third finding is that \texttt{contested\_merge} remains unsolved (as shown in Figure~\ref{fig:merge_issue}). All six LLMs cluster between 0.50 and 0.56 in success, with overlapping confidence intervals, and all remain far below the expert. Even the \texttt{always\_proceed} baseline reaches 0.64, which means many LLMs are not merely below expert behavior; they are below a crude commitment prior. The likely missing capability is the specific tactic used by the scripted expert: slow down, infer that the other vehicle is committed, and merge behind it cleanly. Current models tend to either hesitate, over-yield, or express the correct social idea without reliably executing the maneuver. This is where the benchmark stops being a language test and starts exposing the cost of embodied timing.

The fourth finding is that the benchmark is far from saturated. No LLM reaches the expert reference, and the best average score is only 0.68 (Figure~\ref{fig:ranked_mean}). This gap is not a small residual error at the top of a leaderboard; it is the main result. The scenarios require policies that condition on another agent's evolving behavior, not just policies that sound safe or socially aware. The current generation of models can sometimes infer the right latent state, but their performance remains brittle across interaction types.

\begin{figure}[hbt!]
    \centering
    \includegraphics[width=\linewidth]{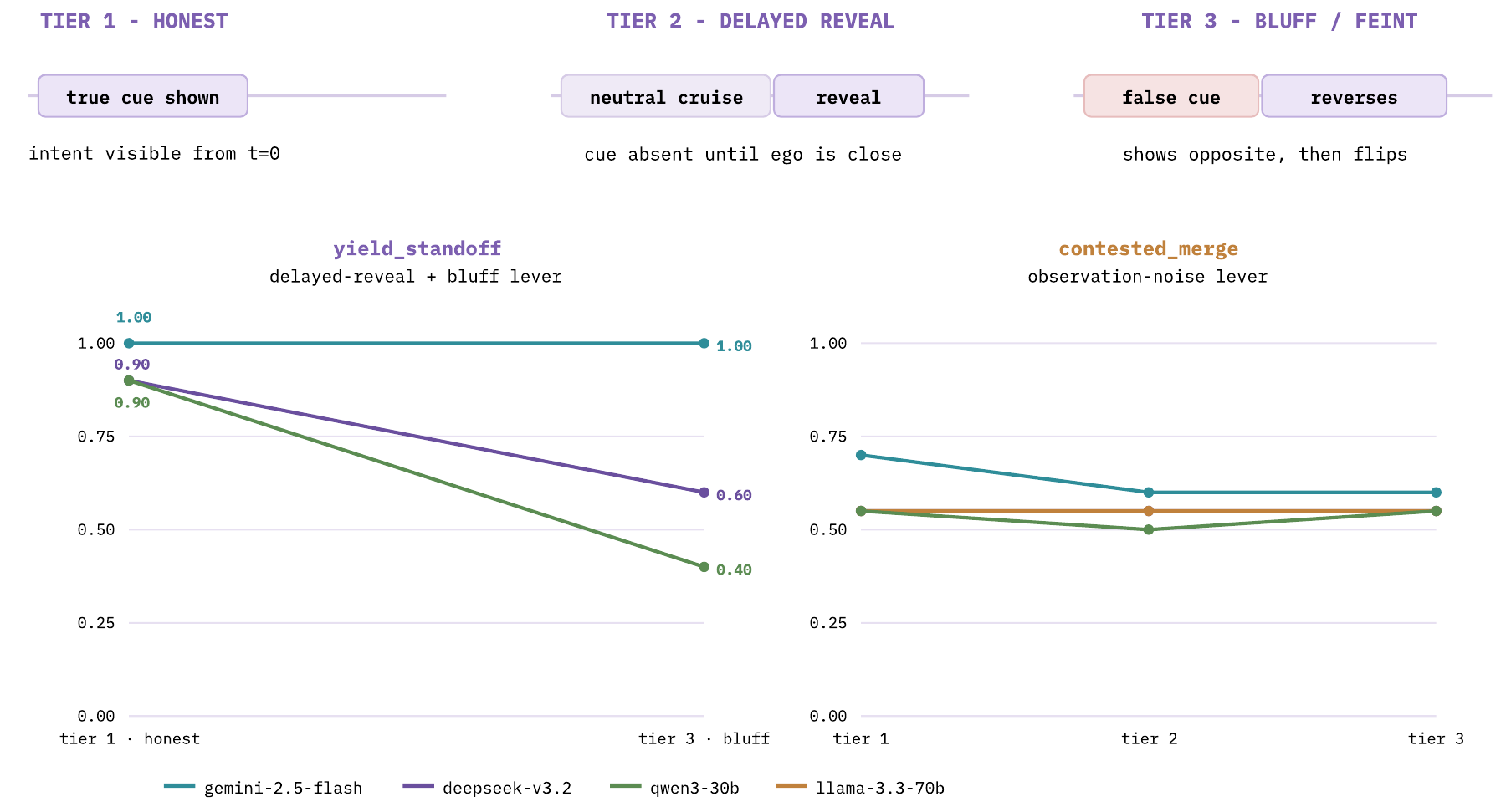}
    \caption{Difficulty tiers stress theory of mind, not perception, and reveal which models hedge vs. follow cues. The \texttt{yield\_standoff} ladder delays the intent reveal (tier 2) and then shows a false cue that reverses when the agent takes the bait (tier 3), so acting on an instantaneous cue becomes unsafe. The bluff de-saturates the scenario exactly as theory of mind predicts: cue-followers collapse (Qwen 0.90$\rightarrow$0.40, DeepSeek 0.90$\rightarrow$0.60) while a model that already waits for a committed signal holds (Gemini $\approx$1.00). \texttt{contested\_merge} keeps the observation-noise lever and stays roughly flat; \texttt{pedestrian\_feint} now has a delayed-step-out tier (tier-1 shown here).}
    \label{fig:bluff_tier_analysis}
\end{figure}

Finally, the bluff-tier analysis (Figure~\ref{fig:bluff_tier_analysis}) confirms that the benchmark distinguishes cue-following from intent reasoning. At the hardest tier, where the relevant signal is delayed and the other agent may feint, cue-following models collapse: \texttt{deepseek-V3.2} drops from 0.90 to 0.60, and \texttt{qwen3-30b-a3b-instruct} drops from 0.90 to 0.40. In contrast, \texttt{gemini-2.5-flash} remains near 1.00, suggesting that it waits for a committed signal rather than reacting to the instantaneous cue. This is the core theory-of-mind distinction the benchmark is meant to expose. A model that reacts to the first visible motion can look competent in easy cases; a model that waits for commitment is doing the harder thing.

\begin{figure}[hbt!]
    \centering
    \includegraphics[width=\linewidth]{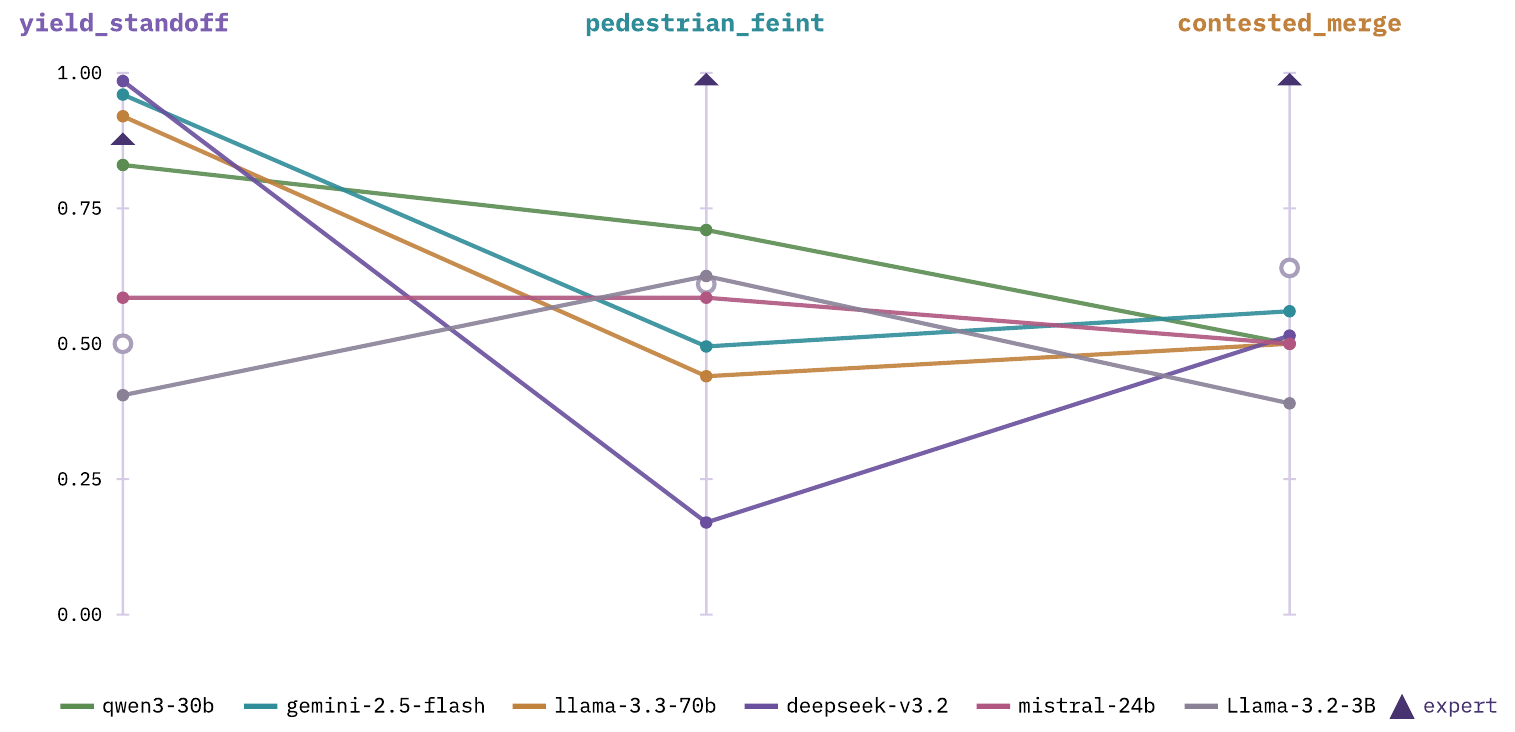}
    \caption{Skill is scenario-specific: the lines cross, so the top-average model loses on two of three scenarios. Parallel coordinates of per-scenario success. Qwen wins the mean but is beaten by DeepSeek and Gemini on \texttt{yield\_standoff}; Gemini leads the merge. The 3B model is mid-pack on pedestrians despite being last overall. Expert ($\Delta$) and floor ($\circ$) are marked on each axis. The crossing pattern argues for reporting per-scenario, not a single headline number.}
    \label{fig:no_winner}
\end{figure}

Overall, the results separate three abilities that are often conflated in language-only evaluations: inferring another agent's intent, selecting the correct maneuver, and executing that maneuver consistently over time. The best models show pieces of this stack, but none yet assemble it reliably (Figure~\ref{fig:no_winner}). They can read the room; the unresolved question is whether they can drive like they understood it.

\section{Fancy a place in the benchmarking paradigm?}

This project aims to address a far more important aspect of self-driving safety, so I don't want it to be just a driving evaluation benchmark. Here are some positional arguments:

\begin{itemize}
    \item It assesses a \textbf{distinct capability} (embodied, real-time, implicit social coordinating under hidden latent) that is absent from explicit bargaining-game benchmarks and perception/VQA AV standards.
    \item It is both \textbf{hard to game} and \textbf{verifiable}. Rather than the model's text prose output, the reward originates from a privileged simulator state and realized actions. The type of “grade the behavior, not the explanation” signal that the reasoning-evaluation literature increasingly requires is \texttt{decision\_correctness}.
    \item By design, it is \textbf{interactive}. The task cannot be reduced to a static dataset since the opponent responds. This is not an annoyance, but a property.
    \item It is resistant to both \textbf{contamination} and \textbf{saturation}. The complexity stages (delayed revelation, bluff) offer a ladder that already de-saturates frontier models, and procedural creation eliminates the need for a preset test set to memorize.
    \item In theory, it is \textbf{RL-ready}. This work contends that the ideal form for RL-based training is a bounded, verifiable, dense-enough reward with a curriculum (tiers) and an expert reference, but it should first prove out as an evaluation.
\end{itemize}

\section{Limitations and Future Work}
\label{sec:limitations}

This benchmark's primary source of control and limitation is its deliberate narrowness. The environment is text-based, meaning that instead of pixels, LiDAR, or recorded sensor streams, the model receives symbolic natural language observations. Perception, tracking, prediction, planning, and control are all necessary for a true driving stack; in this instance, perception is largely taken into account. As a result, the benchmark shouldn't be regarded as a comprehensive evaluation of autonomous driving. It's more akin to a controlled experiment with one missing layer: \textbf{is it possible for a model to monitor and respond to an agent's hidden intent over time?}

This abstraction in the benchmark design is deliberate. Although vision-based driving standards are more realistic in terms of deployment, they also involve a number of failure scenarios, including poor object detection, occlusion handling, motion forecasting, geometry estimation, map priors, control latency, and social reasoning. That is not always diagnostic, but it is realistic. This work eliminates enough perceptual noise by making the scene text-based so that we can pose a more pointed question: \textbf{does the model apply the pertinent behavioral cues once they become available?} So far, the response is only \textit{occasional}. Therefore, the benchmark isolates a tiny but significant cognitive portion of embodied driving evaluation rather than replacing it. It can be compared to testing the negotiation circuit prior to integrating it into the entire robot.

This agentic environment can be effectively formalized as a tiny, partially observed Markov decision process (POMDP). The model only gets sequential cues that update a belief over latent disposition; the other agent's purpose is never directly observed. The cue is early and stable in simple situations. In more difficult situations, it is purposefully false, unclear, or delayed. Maintaining a belief state, obtaining facts through careful inquiry, and committing only after the belief is resolved are all components of a competent policy. This view is seen in Figure~\ref{fig:belief_state}.

Additionally, this POMDP framing suggests a logical extension: environment-based world modeling. The current benchmark assesses the final action sequence and determines if the inferred maneuver corresponds with the hidden disposition using \texttt{decision\_correctness}. A more robust version might require models to forecast future observations under each hypothesis, maintain an explicit belief state over potential intents, and update those beliefs in response to new information. This would change the benchmark from ``select the right action" to ``prove that you are carrying the right world model while acting." This distinction is important because driving does not provide many rewards when a model takes the right action for the wrong reason.

Additionally, the diagnostic intensity of the current circumstances varies. Although \texttt{contested\_merge} discriminates poorly among existing models, it is solvable and not clearly gamed. With overlapping confidence intervals, all six LLMs cluster around 0.50-0.56 success at $n=200$. This indicates that it is difficult to determine which model is superior. Observation noise, a less effective intervention than a pure intent bluff, is its primary difficulty lever. Because the merging geometry contains a "ahead vs. behind" structure, the yield-style feint does not translate cleanly; a merge-ahead strategy can occasionally avoid the bluff rather than resolve it. Therefore, the single highest-value next step is to redesign the merging bluff. The expert strategy should be the target behavior: assume that the other car is committed, decelerate, and merge behind it cleanly.

Although its current tiering is relatively modest, the pedestrian scenario is more diagnostic. Although it is still a subtle bluff, the delayed step-out is a real lever that already distinguishes decisive models from too cautious ones. A \textit{pause-then-dart} pattern, the pedestrian equivalent of the yield feint, would be a more effective pedestrian version. Although conceptually simple, such design has not yet been put into practice. Instead of just responding to the initial motion cue or, worse, freezing indefinitely because a pedestrian is present somewhere in the paragraph, it would assess if the model waits for a committed pedestrian signal.

RL should be handled cautiously, but the payoff is appropriate for assessment. Although the scored reward is verifiable, bounded, and passes the current anti-gaming suite, it is not yet suitable for aggressive optimization. Training rewards and evaluation awards have separate lifetimes. Once an RL policy begins looking for weaknesses, a reward that performs well under passive benchmarking may nevertheless take up odd hobbies. Establishing the evaluation story is therefore the top task right now. Only until the reward passes adversarial optimization tests, such as reward-hacking regressions on trained checkpoints, should RL training begin.

For the primary assertions, the numbers are adequate, but not for the detailed ranks. The benchmark divides the large behavioral clusters: expert versus LLMs, strong yield rules versus weak ones, and the flat merge regime, with $n=200$ rollouts per cell. Larger runs would be necessary to resolve close model pairs, like \texttt{gemini-2.5-flash} versus \texttt{llama-3.3-70b-instruct} on \texttt{yield\_standoff}. Future iterations of the leaderboard should include several rollouts, for example, at least $n \geq 500$ rollouts per cell. It is not necessary to litigate every decimal place, as it is a Formula 1 qualification lap; the current setting is sufficient to see the shape of the failure.

Lastly, the scope is restricted to longitudinal, two-agent, text-only negotiation. Dense multi-agent coordination, explicit vehicle-to-vehicle communication, mixed human-agent traffic, real logged trajectories, and closed-loop vision-language-action control are not covered by the present benchmark. These are logical extensions, but rather than being combined into the initial benchmark, they ought to be added as distinct layers. The goal of the current setting is to keep the first question clear: \textbf{is it possible for a model to deduce hidden intent from sequential behavior and take action based on that belief?} The response is not yet trustworthy. That failure is both big enough to be significant and tiny enough to be studied.

\section{Reproducibility}

The environment is public on the Prime Intellect Environments Hub under owner
\texttt{ashu1069} and environment name \texttt{self-driving-negotiator} (\url{https://app.primeintellect.ai/dashboard/environments/ashu1069/self-driving-negotiator}).

\textbf{Local reproduction}:

\begin{verbatim}
uv pip install -e .
python -m pytest
python scripts/generate_baseline_report.py --episodes 200
python scripts/run_llm_eval.py --model <model> --scenario yield_standoff -n 200 \
  --api-key-var <KEY> --base-url <url>
python scripts/consolidate_llm_results.py
python scripts/plot_radar.py --scenario yield_standoff
python scripts/plot_trajectories.py
\end{verbatim}

\section*{Acknowledgments}

This environment builds on Prime Intellect's \texttt{verifiers} ecosystem and
the Environments Hub model for installable, reproducible LLM environments.

\bibliographystyle{plain}
\bibliography{references}

\end{document}